\documentclass[twocolumn,superscriptaddress,secnumarabic,amssymb, nobibnotes, aps, bm,prl,floatfix,longbibliography]{revtex4-2}
\usepackage[english]{babel}
\usepackage[latin9]{inputenc}
\usepackage[usenames,dvipsnames]{xcolor}
\usepackage{mathtools}
\usepackage{amsmath}
\definecolor{goodgreen}{rgb}{0.1,0.5,0}
\definecolor{goodred}{rgb}{0.7,0,0}
\usepackage[colorlinks,urlcolor=goodgreen,citecolor=blue,linkcolor=goodred]{hyperref}
\usepackage{cleveref}
\makeatletter

\usepackage{soul}

\usepackage{braket}

\usepackage[normalem]{ulem}

\newcommand{\beq}{\begin{equation}}
\newcommand{\eeq}{\end{equation}}
\newcommand{\bea}{\begin{eqnarray}}
\newcommand\bal{\begin{aligned}}
\newcommand\eal{\end{aligned}}
\newcommand{\eea}{\end{eqnarray}}

\makeatother
\begin{document}
\title{Boundary Witten effect in multi-axion insulators}

\author{Giandomenico Palumbo}
\email{giandomenico.palumbo@gmail.com}
\affiliation{School of Theoretical Physics, Dublin Institute for Advanced Studies, 10 Burlington Road, Dublin D04 C932, Ireland}

\begin{abstract}
\noindent 

We explore novel topological responses and axion-like phenomena in three-dimensional insulating systems with spacetime-dependent mass terms encoding domain walls. Via a dimensional-reduction approach, we derive a new axion-electromagnetic coupling term involving three axion fields. This term yields a topological current in the bulk and, under specific conditions of the axions, real-space topological defects such as magnetic-like monopoles and hopfions. Moreover, once one the axions acquires a constant value, a nontrivial boundary theory realizes a (2+1)-dimensional analog of the Witten effect, which shows that point-like vortices on the gapped boundary of the system acquire half-integer electric charge. Our findings reveal rich topological structures emerging from multi-axion theories, suggesting new avenues in the study of topological phases and defects.

\end{abstract}
\date{\today}
\maketitle

\section{Introduction}

Topological phases of matter exhibit remarkable electromagnetic and gravitational responses, often encoded in effective field theories with topological terms \cite{Qi2,Ryu,Fradkin2013,Palumbo,Palumbo11,Tiwari,Palumbo18,Palumbo2, Witten3, Palumbo2013,Palumbo3,Palumbo9,Arouca2022,Finch}. A notable example is the axion electrodynamics in 3+1 dimensions ((3+1)-D) \cite{Wilczek,Coriano3}, arising in the low-energy description of topological insulators \cite{Qi,Essin,Armitage,Nomura} and axion insulators \cite{Vanderbilt,Vanderbilt2,Jo,Wieder,Wieder2,Devescovi}. This theory can exhibit exotic phenomena such as the Witten effect \cite{Witten1979,Franz2}, where magnetic monopoles \cite{Weinberg} acquire half-integer electric charge in the presence of an axion term. Importantly, besides the higher-energy-physics framework \cite{Peccei,Weinberg2,Sikivie}, dynamical axions in solid-state physics have been theoretically studied in several works \cite{Zhang2010,Wang2013,You,Mottola} and their experimental existence has been recently confirmed in quantum materials \cite{Qiu}. Thus, there is currently a strong motivation to figure out if other realistic quantum systems can support more exotic axion physics such as multiple axions, which have been already introduced and studied in high energy physics  \cite{Peloso,Coriano1,Coriano2,Dimopoulos,Montero,Fraser}, but largely overlooked in condensed matter physics with a few exceptions \cite{Qi,Qi3,Lin,Bertolini2025}.\\
In this work, we will investigate a general framework where three appear novel axion-electromagnetic (AX-EM) coupling terms induced by the presence of multiple axion fields. At microscopic level, we will show that these pseudoscalar fields arise in insulating systems that we dub \emph{multi-axion insulators}, which are characterized by the presence of multiple space-dependent mass terms in the effective Dirac Hamiltonian encoding domain walls.  Specifically, we will derive the novel AX-EM coupling  performing a triple dimensional reduction of a (6+1)-dimensional Chern-Simons theory (see, Figure 1) and show that suitable configurations of the axion fields in the bulk states can give rise to real-space topological defects such as magnetic-like monopoles \cite{Weinberg} and hopfions \cite{Sutcliffe,Kuznetsov,Ranada,Hoyos,Trueba}. Moreover, once one of the axion fields acquire a constant value, a non-trivial boundary theory emerges supporting a (2+1)-dimensional version of the Witten effect. Through this novel quantum effect, we will show that vortices living on the gapped boundary of the system acquire half-integer electric charge, which represents an unique feature of our system.
Our results reveal how multi-axion theories go beyond the standard axion insulators by supporting intriguing topological bulk defects and novel boundary effects that would be impossible to obtain in quantum systems with a single axion field.
\begin{figure}[tbph]
	\centering
	\includegraphics[width=8.65cm]{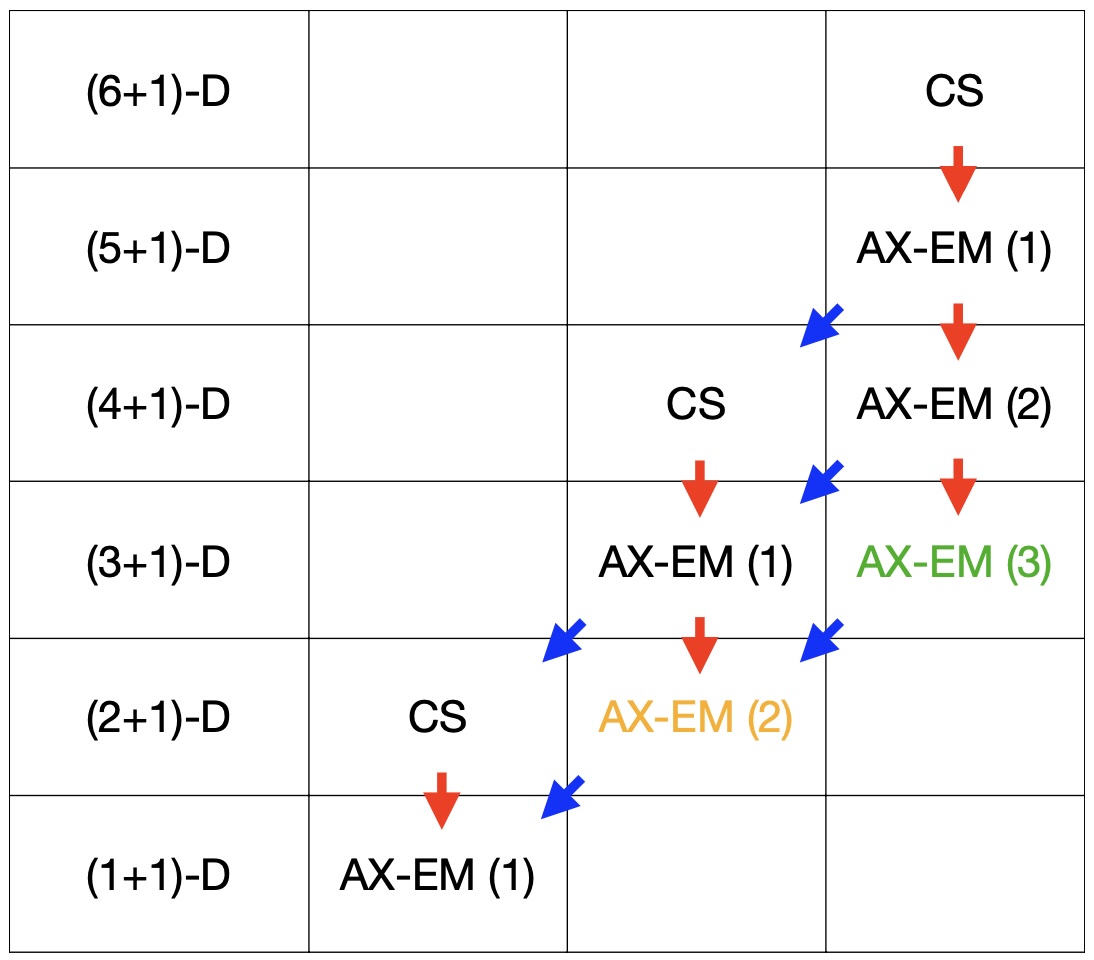}
	\caption{Dimensional reduction of Chern-Simons theories (CS) that give rise to axion-electromagnetic coupling terms (AX-EM (n)). Here, n indicates the number of axion fields while the red arrows represent the dimensional reduction via compactification. Blu arrows refer to the relation between two theories once a single axion field acquires a constant value in the higher-dimensional theory such that the corresponding lower-dimensional action lives on the boundary of the parent theory via the Stokes' theorem. In this work, we mainly focus on the AX-EM theories coloured in green and yellow.}
	\label{}
\end{figure}

\section{Dimensional Reduction and Multi-Axion Terms}

In this section, we will derive an effective field theory that naturally generalizes the standard axion electrodynamics associated with 3D axion insulators \cite{Vanderbilt,Vanderbilt2,Jo,Wieder,Wieder2,Devescovi}. As a warm up, we first discuss the case of convention AXI-EM coupling in both 1D and 3D that involve a single pseudoscalar field \( \theta(x) \), which couples to the electromagnetic field $A_{\mu}$.
The (1+1)-D AXI-EM coupling is given by the following Goldstone-Wilczek action \cite{Goldstone}
\begin{equation}
	S_{\text{GW}}= \frac{1}{4\pi} \int d^2x\, \epsilon^{\mu\nu} \theta F_{\mu\nu},
\end{equation}
where \( \{\mu,\nu\} = \{0,1\} \) and $\epsilon$ is the Levi-Civita tensor. Here,  the axion field acquires a constant value only by imposing time-reversal symmetry. In this case, the above action represents the effective topological theory of the Su-Schrieffer-Heeger model \cite{Qi}.
This topological term can be derived starting from the (2+1)-D Chern-Simons action
\begin{equation}
	S_{CS}^{2+1} = \frac{C_1}{4\pi} \int d^3x\, \epsilon^{\mu\nu\lambda} A_\mu \partial_\nu A_\lambda,
\end{equation}
via a dimensional reduction, with $C_1$ the first Chern number. In fact, we can compactify this theory on the circular direction $x_2\rightarrow x_2+ 2\pi R$, where $R$ is the radius
such that a axion field is defined as \cite{Reece}
\begin{equation}
	\theta = \int_0^{2\pi R} dx_2\, A_{x_2}, 
\end{equation}
which behaves as a periodic pseudoscalar $\theta\rightarrow \theta+2\pi$. In a more compact notation that we use from now on, we can say that a 
dimensional reduction of $S_{CS}^{2+1}$ with \( A_2 \to \theta \)  yields the Goldstone-Wilczek action $S_{\text{GW}}$. We can generalize this approach to the (3+1)-D case, where we have the following axion term
\begin{equation}
	S_{\theta} = \frac{1}{32\pi^2} \int d^4x\, \theta(x) \epsilon^{\mu\nu\lambda\rho} F_{\mu\nu} F_{\lambda\rho},
\end{equation}
with  \( \{\mu,\nu,\lambda,\rho\} = \{0,1,2,3\} \).
Similarly to the previous case, \( \theta \) is quantized to \( 0 \) or \( \pi \) by time-reversal symmetry $\mathcal{T}$, but is in general space-dependent in dynamical axion insulators that break $\mathcal{T}$ and inversion symmetry.
The above action can be derived by integrating out the relativistic fermions in Dirac models with space-dependent mass. In a complementary way, it can also be derived from the dimensional reduction of a (4+1)-D Chern-Simons theory describing a 4D Chern insulator \cite{Qi}, namely
\begin{eqnarray}
	S^{4+1}_{CS}=\frac{C_2}{32\pi^2} \int d^5x\, \epsilon^{\mu\nu\lambda\delta\sigma}A_\mu F_{\nu\lambda}F_{\delta\sigma},
\end{eqnarray}
with $A_4 \to \theta$,
while $C_2$ is the second Chern number that we can set equal to one for simplicity.\\
We now employ this dimensional reduction approach to derive a novel multi-AXI-EM coupling term. We start considering a (6+1)-D Dirac theory with $8 \times 8$ Dirac matrices $\gamma^\mu$ and perform a triple dimensional reduction that gives rise to an effective topological action in (3+1) dimensions. The (6+1)-D massive Dirac theory is defined by the Lagrangian
\begin{equation}
	\mathcal{L}_{6+1}^D = \bar{\psi} \left(i \gamma^\mu \partial_\mu +  \gamma^\mu A_\mu - m \right) \psi,
\end{equation}
where $\psi$ is the fermion field, $\bar{\psi}=\psi^{\dagger}\gamma^0$, \( \mu = \{0,1,\dots,6\} \), and \( A_\mu \) is the background electromagnetic potential. This theory can be seen as the continuum low-energy limit of a 6D Chern insulator coupled to an electromagnetic field.
Integrating out the massive fermions yields the following Chern-Simons action \cite{Koshino,Qi}
\begin{equation}
	S^{6+1}_{CS} = \frac{C_3}{1152\pi^3} \int d^7x\, \epsilon^{\mu\nu\lambda\rho\delta\zeta\eta} A_\mu F_{\nu\lambda} F_{\rho\delta} F_{\zeta\eta},
\end{equation}
with \( C_3 = \mathrm{sign}(m) \) the third Chern number that we set equal to one. Notice that besides Chern insulators, this topological number also characterizes 6D fractional quantum Hall effect for extended objects \cite{Palumbo14}.
Performing now a triple dimensional reduction, we reinterpret the components \( A_4, A_5, A_6 \) as three scalar fields
\begin{equation}
	A_4 \to \theta_4, \quad A_5 \to \theta_5, \quad A_6 \to \theta_6,
\end{equation}
which can be replaced in the above action generating the following multi-AXI-EM term
\begin{equation}
	S_{M\theta} = \frac{1}{32\pi^3} \int d^4x\, \epsilon^{\mu\nu\lambda\rho} \epsilon^{abc} \theta_a \partial_\mu \theta_b \partial_\nu \theta_c F_{\lambda\rho},
\end{equation}
where \( \{a,b,c\} = \{4,5,6\} \) and the greek indices run from 0 to 3. Thus, differently from the standard AXI-EM term that involves a single scalar field, here we have obtained an effective topological action with three independent pseudoscalar fields. Microscopically, these fields are related to three distinct spacetime-dependent mass terms that can be considered in a (3+1)-D Dirac theory with $8 \times 8$ Dirac matrices. In fact, with this specific reducible representation of the Clifford algebra, there exist seven independent $\gamma$. Four of them can be used to write the kinematic term for the Dirac Lagrangian while the last four ones contribute to the mass terms
\begin{equation}
	\mathcal{L}_{3+1}^D = \bar{\psi} \left( i\gamma^\mu \partial_\mu + \gamma^\mu A_\mu - \gamma^a m_a (x)-m \right) \psi,
\end{equation}
where there is summation on the latin index $a$, although we will consider, for practical reasons, an uniform configuration with $m_4=m_5=m_6\equiv m$. Importantly, $S_{M\theta}$ can be directly derived from the above Dirac theory by integrating out the fermionic degrees of freedom at one loop by generalizing the Callan-Harvey anomaly inflow argument \cite{Callan}. Moreover, this Dirac model can be seen as the continuum limit of a multi-axion insulator. The corresponding tight-binding Hamiltonian in real space is given by
\begin{eqnarray}
	H^{3D}_{M\theta}=\sum_{r,s}\left[ \psi^{\dagger}_r \left(\frac{\gamma^0-i \gamma^s}{2}\right)e^{i A_{x,x+s}}\, \psi_{r+s}+h.c.
\right]+ \nonumber \\
\sum_{r,a} \psi^{\dagger}_r \left[\gamma^a\sin \theta_a+\gamma^0(m+\cos \theta_a) \right] \psi_r, \,\,\,
\end{eqnarray}
where $r$ is the position index that defines cubic lattice sites and \( s = \{1,2,3\} \). By assuming translational symmetry and after applying the Fourier transformation, the corresponding momentum-space kernel Hamiltonian describes an insulating eight-band model and reads
\begin{eqnarray}
	\mathcal{H}^{3D}_{M\theta} (k, \theta_a)=\hspace{3.0cm}\nonumber \\
	\sum_{s,a}\left[ \gamma^s \sin k_s + \gamma^a \sin \theta_a+ \gamma^0(m+\cos k_s+\cos \theta_s)  \right], \,
\end{eqnarray}
where $k$ represents the momentum space coordinates and we have switched off the electromagnetic potential for simplicity.
We remind here that axion insulators, differently from standard topological insulators, support gapped boundary states while their bulk breaks time-reversal symmetry but preserve inversion symmetry such that the axion field acquires a quantized value \cite{Vanderbilt,Vanderbilt2}. However, inversion symmetry can be broken by spatially varying axion fields \cite{Zhang2010,Wang2013} that can be induced by the presence of domain walls separating regions with distinct topological or magnetic orders, enabling electromagnetic responses governed by the anomaly inflow. For instance, the gapless modes trapped by domain walls are stabilized by the Callan-Harvey mechanism \cite{Callan}, where bulk currents from the axion electrodynamics flow into the domain walls, explaining the existence of hinge states in axion insulators. Refocusing now on symmetries, we can see that, similarly to the case of a single axion field, the three axion fields, under time-reversal symmetry,  transform as follows
\begin{equation}
	\mathcal{T}: \theta_a \to -\theta_a \Rightarrow \theta_a = 0, \pi,
\end{equation}
implying that $S_{M\theta}$ vanishes unless time-reversal is broken. This is in agreement with the microscopic features of our multi-axion insulators, in which time-reversional symmetry is broken due, for instance, to magnetization.

\section{Composite BF theory, monopoles and hopfions}

In this section, we want to study some bulk properties of the multi-axion theory introduced in the previous section in relation to 3D topological defects, such as magnetic monopoles \cite{Weinberg} and hopfions \cite{Sutcliffe,Kuznetsov,Ranada,Hoyos,Trueba}.
We start rewriting the multi-AXI-EM coupling term $S_{M\theta}$ as follows
\begin{equation}
	S_{M\theta} = \frac{1}{32\pi^3} \int d^4x\, \epsilon^{\mu\nu\lambda\rho} B_{\mu\nu} F_{\lambda\rho},
\end{equation}
where we have introduced the following antisymmetric tensor
\begin{equation}
	B_{\mu\nu} = \epsilon^{abc} \theta_a \partial_\mu \theta_b \partial_\nu \theta_c,
\end{equation}
which can be seen as a composite Kalb-Ramond field \cite{Kalb} that transforms as $B_{\mu\nu} \rightarrow B_{\mu\nu}+2\pi$. This model resembles a composite version of the Abelian BF theory \cite{Palumbo4}, and can in principle support topological defects that behave as extended objects. After adding a minimal coupling between the electromagnetic field and a charge current $J^{\mu}$, from $S_{M\theta}$, we obtain the following equations of motion
\begin{equation}
	J^\rho = \frac{1}{96\pi^3} \epsilon^{\mu\nu\lambda\rho} H_{\mu\nu\lambda},
\end{equation}
where the gauge-invariant three-form curvature tensor is defined as
\begin{equation}
	H_{\mu\nu\lambda} = \epsilon^{abc} \partial_\mu \theta_a \partial_\nu \theta_b \partial_\lambda \theta_c.
\end{equation}
We now consider $\theta_4 = t$, such that the mixed spacetime components simplify
\begin{equation}
	H_{ij t} = \epsilon^{ab} \partial_i \theta_a \partial_j \theta_b\equiv \mathcal{F}_{ij},
\end{equation}
where $\{a,b\}=\{5,6\}$ and $\mathcal{F}_{ij}$ can be seen as a composite electromagnetic field strength. Because we are interested in 3D space-like defects, we can consider the following expressions for $\theta_5$ and $\theta_6$
\begin{equation}
	\theta_5={\rm ArcTan}\left(\frac{x_2}{x_1}\right) , \hspace{0.7cm} \theta_6=\frac{x_3}{\sqrt{x_1^2+x_2^2+x_3^2}},
\end{equation}
such that
\begin{equation}
	\epsilon^{ijk}\mathcal{F}_{ij}=\frac{x^k}{(x_1^2+x_2^2+x_3^2)^{3/2}}\equiv \mathcal{B}^k,
		\end{equation}
coincides with the magnetic field $\mathcal{B}$ of a magnetic monopole.
In this way, we can calculate the topological charge of the monopole
\begin{equation}
	C_1=\frac{1}{2\pi}\int_{S^2} dS_k\mathcal{B}^k,
\end{equation}
which is the first Chern number in real space while $S^2$ is the two-dimensional sphere that wraps the point-like monopole. Thus, we have shown that there exists a suitable configuration for the three axion fields that generates an effective magnetic-like monopole in our system.
In this case, it is possible to define an effective scalar potential $\phi$ such that
\begin{equation}
	\mathcal{B}^k = \partial^k \Phi, \hspace{0.7cm} \Phi= 1/\sqrt{(x_1^2+x_2^2+x_3^2)},
\end{equation}
to obtain a simplified expression for the current
\begin{equation}
	J^k = \frac{1}{96\pi^3} \partial^k \Phi.
\end{equation}
We now discuss more general configurations for the axion fields to induce extended topological solitons known as hopfions \cite{Sutcliffe}, which appear, for instance, as vortex filaments and rings in fluidodynamics \cite{Kuznetsov} and as knotted solutions of the electric and magnetic fields in vacuum \cite{Ranada}. Due to the fact that $\theta_5$ and $\theta_6$ formally resemble real Clebsch potentials, we can consider \cite{Ranada,Hoyos}
\begin{equation}
	\theta_5=\frac{{\rm arg}(\varphi)}{2\pi} , \hspace{0.7cm} \theta_6=\frac{1}{1+|\varphi|^2},
\end{equation}
where 
\begin{equation}
	\varphi=\frac{2(x_1 + i x_3)}{2 x_3 + i (\sqrt{(x_1^2+x_2^2+x_3^2)}-1)},
\end{equation}
which gives rise to a quantized Hopf invariant
\begin{equation}
	Q_H=\int d^3x \mathcal{A}_i \mathcal{B}^i =1,
\end{equation}
with $\mathcal{A}_i=\epsilon^{ab}\theta_a \partial_i \theta_b$ the composite electromagnetic potential. Moreover, further configurations in our system, formally equivalent to electromagnetic knots  \cite{Trueba}, can also be defined. Thus, we have shown that in our 3D multi-axion insulators there exist non-trivial topological configurations of the axions, which uniquely characterized our systems and are absent in standard axion insulators paving the way to unveil the existence of magnetic monopoles and hopfions in novel quantum materials \cite{Rybakov}. Finally we envisage novel possible 4D spacetime textures \cite{Alonso} by allowing all the three pseudoscalar fields to be spacetime dependent to generate, for instance, dynamical tensor monopoles \cite{Palumbo2018,Palumbo6,PalumboScience}. These spacetime-dependent axion fields will be analyzed in detail in a future work.

\section{Boundary Witten effect and vortices}

Here, we consider and study a special case of the multi-AXI-EM term, which becomes a total derivative and induces a (2+1)-D Witten effect on the boundary as we show below.
By fixing \( \theta_4 = \pi \) from the outset, the corresponding reduced multi-axion term reads
\begin{equation}
	S'_{M\theta} = \frac{1}{32\pi^2} \int d^4x\, \epsilon^{\mu\nu\lambda\rho} \epsilon^{ab} \partial_\mu \theta_a \partial_\nu \theta_b F_{\lambda\rho},
\end{equation}
with \( \{a,b\} = \{5,6\} \), and induces the following boundary term
\begin{equation}
	S^{2+1}_{M\theta} = \frac{1}{32\pi^2} \int d^3x\, \epsilon^{\mu\nu\lambda} \epsilon^{ab} \theta_a \partial_\mu \theta_b F_{\nu\lambda},
\end{equation}
which coincides with the effective theory already discussed in the context of charge teleportation \cite{Yonekura} and quantum spin Hall insulators \cite{Qi}.
Here, we are going to show that this theory, which describes the gapped states of certain multi-axion insulators, hosts a (2+1)-D version of the Witten effect that we dub \emph{boundary Witten effect}. Differently, from the (3+1)-D Witten effect that involves 3D magnetic monopoles, in our lower-dimensional case, vortices will acquire a half-integer electric charge via the axion coupling.
Let us first introduce the kinetic term for the electromagnetic field together with its coupling with external sources. The full action reads
\begin{align}
	S^{2+1} &= \int d^3x \bigg[ \frac{1}{16\pi^2} \epsilon^{\mu\nu\lambda} (\partial_\mu \theta_5 \partial_\nu \theta_6 - \partial_\nu \theta_5 \partial_\mu \theta_6) A_\lambda \nonumber  \\
	&\quad - \frac{1}{4} F_{\mu\nu} F^{\mu\nu}  - J^\mu A_\mu \bigg].
\end{align}
By varying this action with respect to the gauge field, we obtain the following equations of motion
\begin{equation}
	\partial_\mu F^{\mu\nu} + \frac{1}{16\pi^2} \epsilon^{\mu\nu\lambda} (\partial_\mu \theta_5 \partial_\lambda \theta_6- \partial_\lambda \theta_5 \partial_\mu \theta_6) = J^\nu,
\end{equation}
that we can also rewrite in a non-covariant form
\begin{eqnarray}\label{noncovariant}
	\partial_i E^{i}=\rho, \nonumber \\
	-\partial_0 E^i+\epsilon^{ij} \partial_j B+\frac{1}{4 \pi^2} \partial_0 \theta_5 \tilde{B}^i=J^i,
\end{eqnarray}
where we have assumed for simplicity $\theta_5=\theta_5(t)$. Here, $E^i=F^{i0}$ is the electric field and
\begin{eqnarray}
	B=\frac{1}{2}\epsilon_{ij} F^{ij}, \hspace{0.3cm} \tilde{B}^i=\frac{1}{2}\epsilon^{ij} \partial_{j}\theta_6,
\end{eqnarray}
are the magnetic fields for $F$ and $d \theta_6$, respectively. By taking the divergence of the second expression in Eq. (\ref{noncovariant}) and assuming $J^i=0$, we have
\begin{eqnarray}
	-\partial_0\partial_i E^i
	+ \frac{1}{4 \pi^2} \partial_0 \theta_5 \partial_i \tilde{B}^i=0,
\end{eqnarray}
which, due to the first expression in Eq. (\ref{noncovariant}), can be rewritten as follows
\begin{eqnarray} \label{Witten23}
	\partial_0 \rho=
	\frac{1}{4 \pi^2} \partial_0 \theta_5 \partial_i \tilde{B}^i.
\end{eqnarray}
We now introduce point-like vortices in the system such that
\begin{eqnarray}
	\partial_i \tilde{B}^i = w_{1D} \delta^2(x),
\end{eqnarray}
where $\delta^2(x)$ is the two-dimensional Dirac delta function and $w_{1D}$ is the topological winding number of the vortices. Notice that this vortex configuration is generated by the following effective magnetic field
\begin{eqnarray}
	\tilde{B}^i= \frac{w_{1D} x^i}{|x|^2}.
\end{eqnarray}
Moreover, we remind that the definition of electric charge reads
\begin{eqnarray}
	Q_e= \int d^2 x\, \rho.
\end{eqnarray}
By integrating Eq.(\ref{Witten23}) over space and time, we can relate the electric charge with the axion fields as follows
\begin{eqnarray}
	Q_e= 
	\frac{1}{2 \pi} \Delta \theta_5\, w_{1D},
\end{eqnarray}
where $\Delta$ represents the net change related to $\theta_5$. If we assume that there
was no initial electrical charge bound to the vortex then $\Delta \theta_5=\{0,\pi\}$,
such that the above equation gives us
\begin{eqnarray}
	Q_e= \frac{1}{2}  w_{1D}.
\end{eqnarray}
Thus, we have shown that point-like vortices acquire a half-integer electric charge through the axion fields. Importantly, these results are in agreement with the fermionic picture, in which the gapped boundary is microscopically described by a massive 2D Dirac Hamiltonian and vortices have been shown to acquire a semi-integer electric charge \cite{Mudry,Franz,Furusaki,Aoki}.
However, differently from these previous works concerning 2D topological systems, our results about the 2D gapped boundary are completely linked to the 3D gapped bulk states and fully encoded in the corresponding multi-axion theory.

\section{Conclusion and Outlook}

Summarizing, in this work, we have introduced a theoretical framework for multi-axion insulators, a new class of quantum systems characterized by the interplay of multiple axion fields coupled to an external electromagnetic potential. These multi-axions, emerging from spatially varying mass terms in an effective Dirac Hamiltonian, give rise to a rich structure of topological phenomena both in the bulk and on the boundary of the system. In particular, our analysis shows that multi-axion configurations can support exotic bulk topological defects, including monopole-like defects and hopfions, which do not have analogues in conventional single-axion systems. Furthermore, we have uncovered a novel boundary effect, namely a (2+1)-D Witten effect, where vortices acquire fractional electric charge due to the presence of the axion fields.
These findings open several promising directions for future research. The classification of multi-axion phases in terms of their topological invariants and defect structures also deserves further investigation. Future work may explore generalizations of multi-axion theories involving axial gauge fields induced by strain \cite{Wieder3,Cortijo3}, non-Abelian gauge fields related to spin \cite{Ludwig,Lin2}, axion-gravitational coupling generated by a curved background geometry \cite{Schupp,Melle}, and possible implications for quantum anomalies \cite{Arouca2022} and dualities \cite{Palumbo4} in both condensed matter and high-energy contexts. On the experimental side, our results suggest new avenues for realizing and probing multi-axion physics in synthetic quantum systems, such as cold atoms \cite{Bermudez,Xie} and topological photonic crystals \cite{Devescovi2}, where multiple spatially modulated mass terms may be tunable.\\
Overall, our work highlights how generalizations of axion electrodynamics to systems with multiple axions can lead to fundamentally new phenomena in topological phases of matter, motivating further exploration of their theoretical foundations and experimental realizations.

\noindent

%\vspace{0.2cm}

\noindent {\bf Acknowledgements:} We thank Claudio Corian\'o for useful discussions.

\bibliography{references}

\end{document}